\documentclass{aa}

\usepackage{color}

\usepackage{amsmath}
\usepackage{float}
\usepackage{txfonts}

\usepackage{hyphsubst}
\HyphSubstLet{english}{usenglishmax}
\usepackage[english]{babel}

\usepackage{hyperref}
\usepackage{booktabs,tabularx}
\usepackage{adjustbox}

\usepackage[flushleft]{threeparttable}

\usepackage{comment}
\usepackage{graphicx}
\usepackage{natbib}
\usepackage{tablefootnote}

\usepackage[nodisplayskipstretch]{setspace}
\usepackage{multirow}

\usepackage[usenames,dvipsnames]{xcolor}
\usepackage{soul}

\begin{document}

\title{Hard X-ray irradiation of cosmic silicate analogs: \\ structural evolution and astrophysical implications}

\author{L. Gavilan\inst{1} 
          \and
         C. J\"ager \inst{2}
          \and
         A. Simionovici \inst{3}
          \and
          J. L. Lemaire \inst{4}
         \and 
          T. Sabri\inst{2}
         \and
          E. Foy \inst{5}
	 \and
          S. Yagoubi \inst{6}
          \and
         T. Henning \inst{7}
          \and
          D. Salomon \inst{8}
          \and
         G. Martinez-Criado \inst{8}
          }

\institute{Institut d\textsc{\char13}Astrophysique Spatiale (IAS), CNRS,  Univ. Paris Sud, Universit\'e Paris-Saclay, F-91405 Orsay, France
         \and
         Laboratory Astrophysics and Cluster Physics Group of the Max Planck Institute for Astronomy at the Friedrich Schiller University \&
Institute of Solid State Physics, Helmholtzweg 3, 07743 Jena, Germany
          \and
          Institut des Sciences de la Terre, Observatoire des Sciences de l'Univers de Grenoble, BP 53, 38041 Grenoble, France
          \and
         Institut des Sciences Mol\'eculaires d'Orsay (ISMO), CNRS, Univ. Paris Sud, Universit\'e Paris-Saclay, F-91405 Orsay, France
           \and
            LAPA-IRAMAT, NIMBE, CEA, CNRS, Universit\'e Paris-Saclay, CEA Saclay, 91191 Gif sur Yvette, France
           \and
            LEEL, NIMBE, CEA, CNRS, Universit\'e Paris-Saclay, CEA Saclay, 91191 Gif sur Yvette, France 
            \and
          Max Planck Institute for Astronomy K\"onigstuhl 17, D-69117 Heidelberg, Germany
          \and
         European Synchrotron Radiation Facility, 71 Avenue des Martyrs, 38000 Grenoble, France\\
             \email{lisseth.gavilan@ias.u-psud.fr}
           }

\abstract {Protoplanetary disks, interstellar clouds, and active galactic nuclei, contain X-ray dominated regions. X-rays interact with the dust and gas present in such environments. While a few laboratory X-ray irradiation experiments have been performed on ices, X-ray irradiation experiments on bare cosmic dust analogs have been scarce up to now.} 
{Our goal is to study the effects of hard X-rays on cosmic dust analogs via in-situ X-ray diffraction. By using a hard X-ray synchrotron nanobeam, we seek to simulate cumulative X-ray exposure on dust grains during their lifetime in these astrophysical environments, and provide an upper limit on the effect of hard X-rays on dust grain structure.}  
{We prepared enstatite (MgSiO$_3$) nanograins, analogs to cosmic silicates, via the melting-quenching technique. These amorphous grains were then annealed to obtain polycrystalline grains. These were characterized via scanning electron microscopy (SEM) and high-resolution transmission electron microscopy (HRTEM) before irradiation. Powder samples were prepared in X-ray transparent substrates and were irradiated with hard X-rays nanobeams (29.4 keV) provided by beamline ID16B of the European Synchrotron Radiation Facility (Grenoble). X-ray diffraction images were recorded in transmission mode and the ensuing diffractograms were analyzed as a function of the total X-ray exposure time.}
 {We detected the amorphization of polycrystalline silicates embedded in an organic matrix after an accumulated X-ray exposure of 6.4 $\times$ 10$^{27}$ eV cm$^{-2}$. Pure crystalline silicate grains (without resin) did not exhibit amorphization. None of the amorphous silicate samples (pure and embedded in resin) underwent crystallization. We analyzed the evolution of the polycrystalline sample embedded in an organic matrix as a function of X-ray exposure.}
  {Loss of diffraction peak intensity, peak broadening, and the disappearance of discrete spots and arcs, revealed the amorphization of the resin embedded (originally polycrystalline) silicate sample. We explore the astrophysical implications of this laboratory result as an upper-limit to the effect of X-rays on the structure of cosmic silicates.}

\keywords{astrochemistry, ISM: dust, extinction, ISM: evolution, methods: laboratory: solid state, methods: analytical}

\maketitle

\section{Introduction}

X-rays are an important component of the radiation field of young stellar objects (YSOs).
In the pre-main sequence phase, emission is dominated by intense flares, called catastrophic flares  \citep{Feigelson2003, Feigelson2005}, which can reach an X-ray luminosity (L$_X$) in the range of 10$^{32}$ erg s$^{-1}$ within hours. Smaller flares (L$_X$ $\sim$ 10$^{31}$ erg s$^{-1}$) produced in YSOs and in the modern Sun last a few hours and cycle every few days. When these young stars enter the main sequence, the X-ray luminosity is in the range of 10$^{30}$ erg s$^{-1}$ \citep{Shu2001}.  
In young stars, X-rays dominate over extreme ultraviolet (EUV) fluxes up to 100 million years (Myrs), and the X-ray/EUV ratio remains within a factor of two for stars as old as 1 Gyr. For today's Sun this ratio is about 0.25 \citep{Ribas2005}.  Thus it is expected that such energetic X-ray radiation, dominating over UV photons, contributes to the processing of circumstellar materials, mostly illuminated by young T-Tauri type stars.  \cite{Shu2001} studied the melting capabilities of observed hard X-ray flares in protostars, suggesting that vapor-phase condensation of flare-evaporated material can lead to small refractory solids. 
\begin{figure*}[ht]
\begin{center}
\includegraphics[width=180mm]{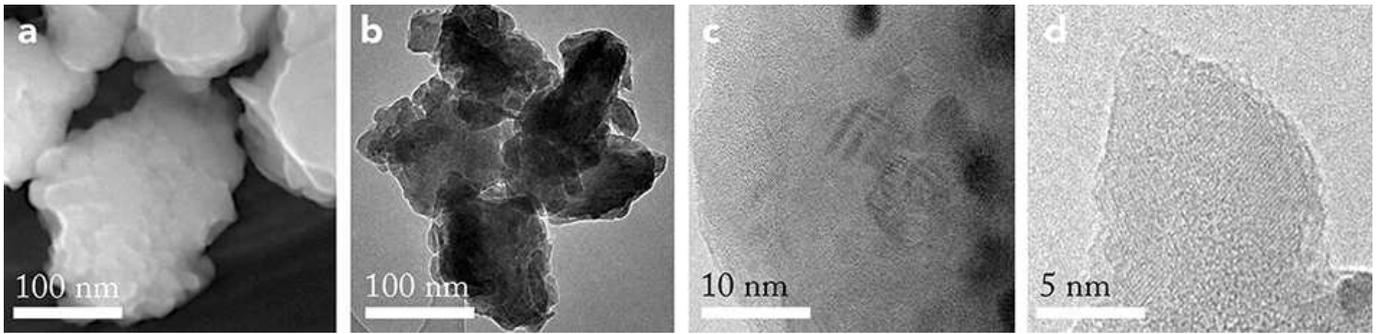}
\caption{MgSiO$_3$ crystals before irradiation prepared by quenching/melting. \textbf{a.} FESEM image showing the agglomerated grains surface, \textbf{b.} HRTEM image showing the stacking of agglomerated grains (100 nm scale),  \textbf{c.} HRTEM image showing crystalline grains in an amorphous matrix (10 nm scale),  \textbf{d.} HRTEM image showing a single crystalline grain (5 nm scale).} 
\label{semtem}
\end{center}
\end{figure*}

Amorphous silicates are abundant in primitive extraterrestrial materials and most astrophysical environments \citep{Henning2010}. The Infrared Space Observatory (ISO) revealed that $>$ 99\% of silicates are amorphous in the interstellar medium (ISM), with a crystalline fraction of 0.2\% $\pm$ 0.2 \% by mass  \citep{Kemper2004}. This is lower than the crystalline fraction observed in the circumstellar envelopes of evolved stars, the main contributors of dust to the ISM. Crystalline silicates are also found in comets and interplanetary dust particles (IDPs) \citep{Molster2005, Nuth2006}. Glass with embedded metals and sulfide (GEMS) in IDPs contain silicates in amorphous and crystalline form, and it was suggested that these crystalline silicates became amorphous by radiation damage \citep{Bradley1994}. Indeed, depending on their astrophysical environment, dust grains can be altered by several energetic mechanisms, both thermal and non-thermal. Thermal annealing, in which silicates are heated up to 1000 K, can crystallize dust grains. Non-thermal sources include photons, electrons, and ions. 
\cite{Molster1999} reported observations of highly crystalline silicate dust in the disks surrounding binary red-giant stars, produced in amorphous form in their outer atmospheres. Since the temperatures in these disks are too low for grains to undergo thermal annealing, they proposed that a non-thermal mechanism is responsible for the crystallization. We propose X-ray irradiation as a possible non-thermal mechanism acting on the structure of cosmic silicates. 

Hard X-rays permeate several astrophysical environments where silicate and other types of dust are present, such as protoplanetary disks surrounding young stars, planetary nebulae, active galactic nuclei, and any X-ray dominated region (XDR). In XDRs, photons can penetrate deeply within icy dust mantles, activating a bulk chemistry competing with surface processes such as UV photons and electron irradiation.  We expect X-rays to have an effect on the structure of dust grains due to thermal transfer and/or charge accumulation during their lifetime in XDRs. 

Few laboratory experiments studying the interaction of X-rays with cosmic dust analogs exist. Dust irradiation experiments have mostly dealt with ices, e.g. \cite{Andrade2010, Ciaravella2012, Chen2013}. Experiments of X-ray irradiation on dust mineral analogs at room temperature are scarce. 
\cite{Simionovici2011a,Simionovici2011b} studied the effect of X-rays on \textit{Stardust} returned dust grains embedded in aerogel, an extremely low density solid ($\rho$ $<$ 0.5 g cm$^{-3}$) used for gently stopping hypervelocity (6 - 10 km s$^{-1}$) grains in space. To estimate the total exposure to X-rays, they considered an upper limit dose that they called the \textit{astrophysical limit} (AL). This is the estimated X-ray exposure on an interstellar dust grain during its lifetime in a diffuse cloud ($\sim$3 $\times$ 10$^7$ years, \cite{McKee1989}). In these clouds, dust grains are exposed to a quasi-isotropic diffuse cosmic X-ray background in the 3-300 keV energy range, most likely produced by active galactic nuclei (AGN) components, as measured by the HEAO 1 spacecraft \citep{Gruber1999}. Considering the X-ray flux integrated over time (fluence) and neglecting other irradiation sources, the estimated X-ray astrophysical limit is 5 $\times$ 10$^{23}$ eV cm$^{-2}$. The \textit{Stardust} grains embedded in aerogel were irradiated using a synchrotron X-ray nanobeam (17 keV, 10$^{10}$ ph s$^{-1}$, 150 x 190 nm) and showed damage effects (radial smearing) mainly attributed to secondary charge accumulation  \citep{Simionovici2011a}. We use the AL as a reference number to compare synchrotron exposure (large flux over a short timescale) to the total exposure in an astronomical time scale (low flux over a large timescale). This exposure time does not consider X-ray damage on dust particles as these processes depend on: integrated time, X-ray energy, elemental composition of dust, and thickness. Since all experiments presented in this paper were done using the same monochromatic X-ray flux we consider the total exposure as the main variable. 

By using high-flux synchrotron X-rays, we seek to understand the role of hard X-rays in the structure and/or morphology of dust grains in astrophysical X-ray dominated regions. We will compare synchrotron fluences to the X-ray exposures on dust particles during their lifetime  in interstellar clouds,  circumstellar, and protoplanetary disks. Because of the nanofocused synchrotron fluxes, we can reach astronomical fluences in less than 1 second (for the ISM) to 45 seconds (for the circumstellar medium, CSM). This paper is organized as follows: in section 2, we describe sample preparation, in section 3, we present the irradiation experiments, in section 4, we describe the results from the X-ray diffraction analysis, in section 5, we discuss the results, in section 6, we present the astrophysical implications of these experiments. Finally, in section 7, we summarize our conclusions. 

\section{Preparation of powder enstatite samples}
Samples were prepared by the Laboratory Astrophysics Group in Jena.  Enstatite (MgSiO$_3$) grains were prepared using the melting-quenching technique \citep{Dorschner1995}. To produce crystalline MgSiO$_3$, magnesium carbonate and silicon dioxide were thoroughly mixed in stoichiometric quantities and filled into a platinum crucible. The powder was heated in a resistance furnace up to a temperature of 1917 K and the melt was kept for one hour at the same temperature. To produce a homogeneous MgSiO$_3$ glass, the melt was poured through rotating copper rolls accomplishing quenching rates of $\sim$1000 K s$^{-1}$. The resulting amorphous glass splats were powdered and finally annealed at a temperature of 1400 K for one hour in air to ensure enstatite formation only. Size  distributions smaller than 500 nm were finally produced by grinding and sedimentation of the annealed crystalline grains in ethanol. The frequently agglomerated grains are characterized by a very broad size distribution ranging from a few tens up to about 400 nm (as seen by FESEM and HRTEM images in Fig. \ref{semtem}: a-b). The individual particles are composed of randomly oriented, nanometer-sized enstatite crystals still embedded in a small fraction of amorphous MgSiO$_3$ matrix that remained after the final annealing step (see HRTEM micrographs Fig. \ref{semtem}: c-d). 

MgSiO$_3$ powder samples were deposited on Si$_3$N$_4$ membranes which were supported by a Si 0.5 $\times$ 0.5 mm frame. The Si$_3$N$_4$ itself is 50 nm thick, and fragile for manipulation. It is also quasi-transparent to X-rays and non-diffracting. In order to insert the powder sample we used a few drops of CHCl$_3$ for dispersion. In addition, to keep the samples in place on the Si$_3$N$_4$ membrane we used cyanoacrylate (C$_6$H$_7$NO$_2$), an acrylic resin that rapidly polymerises in the presence of water (specifically hydroxide ions), forming long, strong chains, joining the bonded surfaces together (commercially known as \textit{superglue}), which is non-diffracting and quasi-transparent to hard X-rays. This had the advantage of keeping the powder samples in place, otherwise repelled from the  Si$_3$N$_4$ film by electrostatic effects. Samples were also deposited in silica capillaries (internal diameter $\phi$ = 780 $\mu$m, and wall thickness = 10 $\mu$m). 

\section{Irradiation with hard X-rays}
\begin{figure}[ht]
\begin{center}
\includegraphics[width=90mm]{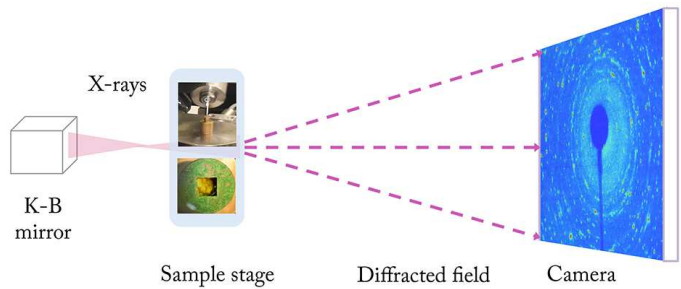}
  \caption{Schema of the powder diffraction setup at the ID16 beamline at the ESRF. X-rays are focused at the sample, consisting on polycrystalline MgSiO$_3$ powder, placed in a capillary ($\phi$ = 780 $\mu$m) or powder deposited into a Si box ($\phi$ = 200 $\mu$m) holding a Si$_3$N$_4$ membrane. The transmission X-ray diffraction image is recorded behind the sample.}
  \label{setup}
\end{center}
\end{figure}
For these experiments we used the recently commissioned ID16B-Nano Analysis beamline at the ESRF synchrotron \citep{MartinezCriado2015}. The 165 m long beamline provides nano-focused beams for analytical imaging. During measurements, the spot size at the focal point is 60 $\times$ 80 nm$^2$. The setup is based on Kirkpatrick-Baez (K-B) multilayer coated Si mirrors, a high-precision piezostage to raster the sample in the beam, and a light microscope to visualize the region of interest on the sample. The setup is schematized in Fig. \ref{setup}. The penetration depth of the 29.4 keV nanobeam ensures that the full sample thickness is irradiated. Based on a double Si (111) crystal monochromator, we used a 29.4 keV beam with a photon flux of 2 - 6 $\times$ 10$^9$ photons s$^{-1}$. While this high energy monochromatic beam does not simulate a spectrally broad stellar X-ray spectrum, it provides an upper limit order of magnitude X-ray exposure on dust particles over their lifetime in astrophysical environments.

We use the integration time during the acquisition of the X-ray diffraction image as the exposure time on dust grains, which allows simultaneous irradiation and recording of the diffraction pattern. In case of continuous irradiation, we could move the diffraction camera (FReLoN CCD) off-sample and once irradiated, return the camera into the beam path and record the resulting diffraction image. 
In order to estimate the optimal sample thickness we used the X-ray database tool (\url{http://henke.lbl.gov/optical_constants}) based on photoabsorption, scattering, and transmission data by \cite{Henke1993}.  For MgSiO$_3$ grains we used a density of 3.189 g/cm$^3$. 
At 29.4 keV, a 200 $\mu$m thick sample has a transmission of 0.95, while a 780 $\mu$m thick sample has a transmission of 0.82. The absorbed energy per photon is 1.47 keV and 5.29 keV respectively. For powder diffraction, 1 $\mu$g to 1 mg of sample powder was required.
\begin{figure*}[ht]
\begin{center}
\includegraphics[width=180mm]{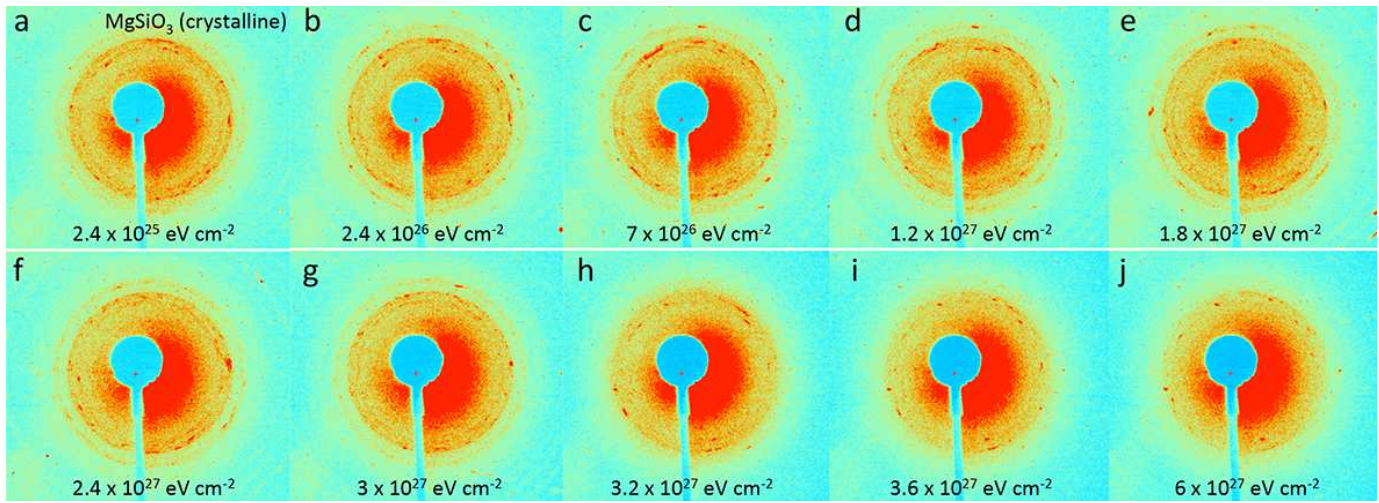}
  \caption{Diffraction patterns for MgSiO$_3$ crystalline powder samples embedded into the Si$_3$N$_4$ substrate at increasing X-ray exposures. \textbf{a.} crystalline MgSiO$_3$, \textbf{b-e.} diffraction rings and arcs visible, \textbf{f-g.} less rings, few arcs and spots, \textbf{h-j.} no rings, a few spots.} 
  \label{encsin}
\end{center}
\end{figure*}
Our initial objective was to irradiate samples until reaching the fluence of 5 $\times$ 10$^{23}$ eV cm$^{-2}$, expected for a dust grain during its lifetime in an interstellar cloud (the AL limit), as it was done for \textit{Stardust} grains embedded in aerogel. With the available synchrotron flux, this was reached after 0.4 - 1.2 seconds. 
We began the irradiation experiments by systematically taking 1 second diffraction images to monitor the effect, and subsequently increased to 10 second images in order to better distinguish any further evolution in the diffraction pattern. 
 We continued irradiating samples when small modifications in the diffraction pattern were detected (attenuation of rings, appearance of arcs, disappearance of spots, etc.), which were used as indicators of  structural modifications at the nanoscale. 

We irradiated the samples beyond the 5 $\times$ 10$^{23}$ eV cm$^{-2}$ dose (AL limit) in order to verify if an accumulated exposure effect could be detected. In a circumstellar disk, a dust grain exposed to a one hour catastrophic flare within the reconnection ring receives an exposure of $\sim$10$^{25}$ eV cm$^{-2}$, and for a typical flare, $\sim$10$^{22}$ eV cm$^{-2}$ \citep{Shu2001}. With the synchrotron nanobeam, these fluences can be achieved after 15 to 45 seconds respectively. If we consider the X-ray luminosity for the TW Hydrae (the closest T-Tauri star to us) integrated from 0.2 to 2 keV, L$_X$ = 2 $\times$ 10$^{30}$ erg s$^{-1}$  \citep{Kastner2002}. At 10 astronomical units (AU), this amounts to a total exposure of $\sim$10$^{26}$ eV cm$^{-2}$ in 1 Myr assuming the X-ray optical depth is $\tau_{X}$ = 1 \citep{Gorti2004, Andrade2010}. An equivalent fluence was achieved under 115 seconds with the synchrotron nanobeam. While nanoparticles are quasi transparent to hard X-rays, soft X-rays will be more easily absorbed. Thus the exposure dose is only an indicator of the integrated incident energy on the grains, not on the adsorbed energy.

We irradiated four types of powder samples of enstatite (MgSiO$_3$) stoichiometry. Sample 1 consisted of crystalline MgSiO$_3$ grains deposited into a resin embedded Si$_3$N$_4$ membrane.  Sample 2 consisted of amorphous MgSiO$_3$ grains deposited into a resin embedded Si$_3$N$_4$ membrane.  Sample 3 consisted of amorphous MgSiO$_3$ grains deposited into a capillary. Sample 4 consisted of crystalline MgSiO$_3$ grains deposited into a capillary. 

\begin{figure}[htb]
\begin{center}
\includegraphics[width=90mm]{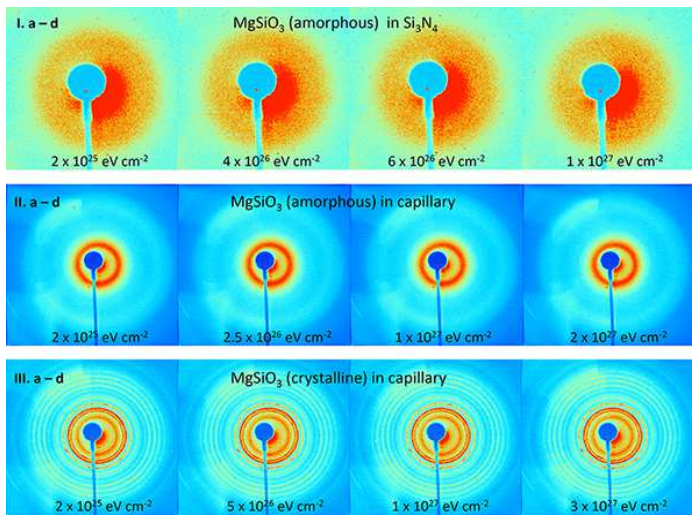}
\caption{Samples showing no evolution in their diffraction pattern with increasing X-ray exposure. \textbf{I.} MgSiO$_3$ amorphous grains in a resin embedded Si$_3$N$_4$ support. \textbf{II.} MgSiO$_3$ amorphous grains in a capillary. \textbf{III.} MgSiO$_3$ crystalline grains in a capillary.}
\label{noamor}
\end{center}
\end{figure}

From these four samples, only sample 1 showed structural changes during irradiation, as seen in Fig. \ref{encsin}. The interpretation of this result will be explored in the next section. The evolution of the initially polycrystalline sample can be seen on the diffraction patterns. The attenuation of diffraction rings reveals an increasing amorphization. The appearance of some spots shows the  preferential orientation due to crystal growth which disappeared under consecutive scans. At an exposure of 3.2 $\times$ 10$^{27}$ eV cm$^{-2}$ (2600 s) the diffraction structure (rings, arcs) is largely attenuated.  After this exposure time, the sample becomes highly amorphized. A detailed analysis of the structural evolution of this sample will be provided in the next section.  
Some of the non-evolving diffraction patterns as a function of X-ray exposure are shown in Fig. \ref{noamor}. 
\section{X-ray Diffraction analysis}
\label{XRD}
\begin{figure}[!ht]
\begin{center}
\includegraphics[width=92mm]{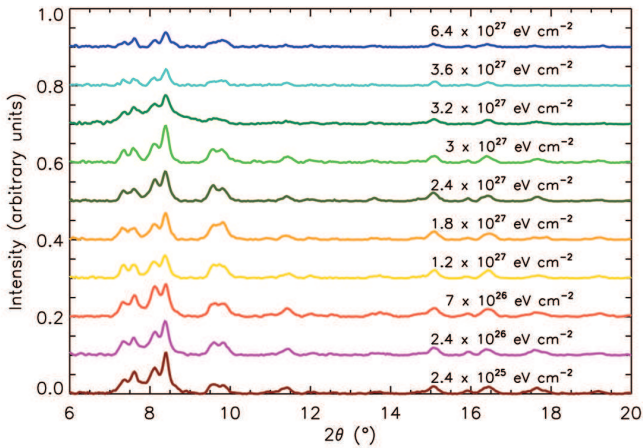}
\caption{Evolution of diffraction peaks of MgSiO$_3$ (clinoenstatite) crystalline grains on a resin-embedded Si$_3$N$_4$ membrane under X-ray exposures from 2.4 $\times$ 10$^{25}$ eV cm$^{-2}$ to 6.4 $\times$ 10$^{27}$ eV cm$^{-2}$. }
\label{xrd10}
\end{center}
\end{figure}
All samples were kept at a fixed position and diffraction images were taken in transmission mode. 
Flat-fielded and dark-subtracted diffraction images were processed with \textit{Fit2D} \citep{Hammersley1996}. The beam center parameters were refined by using the Bragg diffraction rings. The refined wavelength for irradiation at 29.4 keV was found to be $\lambda$ = 0.42105 $\AA$. 

In the rest of this section we will discuss the diffraction analysis as a function of X-ray exposure for sample 1, the MgSiO$_3$ grains embedded in the Si$_3$N$_4$ membrane, as this is the only sample showing  structural changes (Fig. \ref{encsin}). 
The evolution of the diffractograms with increasing X-ray exposure is shown in Fig. \ref{xrd10}. There is a decrease in the intensity of the different peaks as the dose increases in the 2$\theta$ = 9-12$^\circ$ and 15-18$^\circ$ ranges, most clearly seen for the peak at 2$\theta$ = 8.4$^\circ$.  There is also a clear broadening of all peaks starting from 3 $\times$ 10$^{27}$ eV cm$^{-2}$. This reveals an amorphization of the polycrystalline grains falling into the X-ray nanobeam spot. 

\begin{figure}
\begin{center}
\includegraphics[width=86mm]{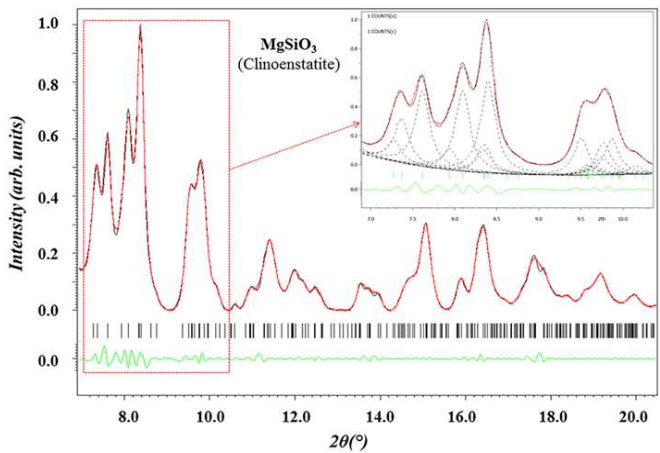}
  \caption{Calculated scattering intensity profile (black) compared to experimental data (red) collected with synchrotron X-ray diffraction of the crystalline MgSiO$_3$ sample (clinoenstatite) embedded in a capillary. }
  \label{lebail}
\end{center}
\end{figure}

The \textit{Jana2006} crystallography software \citep{Petricek2014} for powder analysis, was used for the structural refinement from the 2$\theta$ diffractograms. The peak shapes were adjusted by Lorentzian functions and a ten Legendre polynomial was used for background subtraction. At room temperature and ambient pressure, the measured lattice parameters are in good agreement with the literature values for enstatite reported in the Inorganic Crystal Structure Database (ICSD) \cite{Ohashi1984}. 
\begin{table*}[t]
\centering
\begin{threeparttable}
\caption{Fitted crystal parameters for MgSiO$_3$ grains under X-ray irradiation}
\label{tab1}
\begin{tabular}{cccccccccc}
\toprule
\centering
Exposure  &  Fluence &  \multicolumn{5}{c}{Lattice parameters} &   \multicolumn{2}{c}{Crystal size [ $\AA$ ] } & Fit error   \\
\cmidrule(r){3-7}
\cmidrule(r){8-9}
time [ s ] &  [ eV cm$^{-2}$ ] & a [ $\AA$ ] & b [ $\AA$ ] & c [ $\AA$ ] & $\beta$ [ $^{\circ}$ ] & $\rho$ [ g cm$^{-3}$ ] & D-S$^c$ & W-H$^d$ & R$_p$ \\
\toprule
Ref.$^a$        &                                              & 9.6270 & 8.8340 & 5.1800 & 108.3400        & 3.1890                  &              &         &       \\
20$^{b}$        & 2.42 $\times$ 10$^{25}$      & 9.6445 & 8.8264 & 5.1720 & 108.7902        & 3.1995      &   79.12     & 74.73       & 3.01  \\
20                  & 2.42 $\times$ 10$^{25}$       & 9.6445 & 8.8264 & 5.1720 & 108.7902        & 3.1995       &   77.97   & 72.38       & 9.67  \\
200                & 2.42 $\times$ 10$^{26}$       & 9.6300 & 8.8049 & 5.1542 & 108.6531        & 3.2206      &    86.53     & 89.59       & 11.24 \\
580                & 7.01 $\times$ 10$^{26}$       & 9.6276 & 8.7907 & 5.1641 & 108.7908        & 3.2231      &  76.55    & 76.48       & 9.06  \\
2380              & 2.88 $\times$ 10$^{27}$       & 9.6097 & 8.8269 & 5.1636 & 108.8866        & 3.2180     &  77.97    & 82.91       & 9.36  \\
2670              & 3.23 $\times$ 10$^{27}$       & 9.5725 & 8.8844 & 5.1577 & 109.0414        & 3.2162     &  48.93    & 44.06       & 14.26 \\
\bottomrule
\end{tabular}
\begin{tablenotes}
      \small
      \item $^a$From the Inorganic Crystal Structure Database (ICSD), e.g. \cite{Ohashi1984} for enstatite.
      \item $^{b}$Single sample of MgSiO$_3$ grains in the capillary substrate. All other samples in Si$_3$N$_4$ windows. 
      \item $^c$Debye-Scherrer method.
      \item $^{d}$Williamson-Hall method. 
    \end{tablenotes}
\end{threeparttable}
\end{table*}
The powder X-ray diffraction pattern is unambiguously indexed with the monoclinic space group P2$_1$/c and unit cell parameters a = 9.6445(9) $\AA$, b = 8.8264(8) $\AA$, c = 5.1720(4) $\AA$ and $\beta$ = 108.790(7)$^\circ$. 
The reliability factors after fitting the diffraction peaks (Le Bail refinement, Fig. \ref{lebail}) are R$_p$ = 3.01 and wR$_p$ = 5.35 for the crystalline MgSiO$_3$ sample in the capillary (thickness = 780 $\mu$m, see diffraction patterns II. a-d in Fig. \ref{noamor}). For the crystalline MgSiO$_3$ powder held in a Si$_3$N$_4$ window, the optimized reliability factors are R$_p$ $\sim$10. This is due to the 4 times larger thickness of the capillary sample in comparison to the sample in the Si box, increasing the diffraction contrast. 

The derived crystallite sizes will depend on the instrumental broadening, the monochromator crystals (2 x Si 111), the beam divergence, the focusing (K-B mirrors), the shape of the grains, etc. Since we are more interested in the relative evolution of the crystallite structural properties, we assume that the beam parameters, monochromator, and CCD camera are identical throughout the scans. We estimate the derived error from the fit of all diffraction peaks, which increases naturally due to the decrease in diffraction signal. 
As an inferior limit of the crystallite sizes we use the Debye-Scherrer relation, for which the size is inversely proportional to the full-width at half maximum (FWHM) of the diffraction peaks, i.e.
\begin{equation}
D_v = \frac{k \lambda}{\beta \cdot cos(\theta)}
\end{equation}
where D$_v$ is the crystallite size, $\beta$ is the FWHM in radians, $\theta$ is the diffraction angle and \textit{k} = 0.94 is the Scherrer constant. 

\begin{figure}
\begin{center}
\includegraphics[width=86mm]{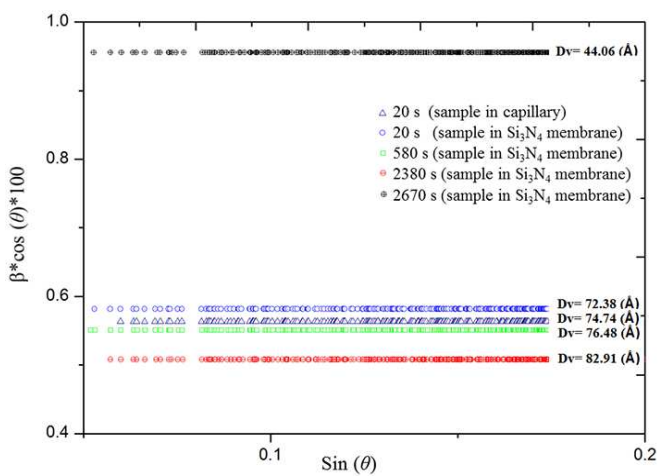}
  \caption{Crystallite size before and during irradiation of the polycrystalline MgSiO$_3$ sample (clinoenstatite) embedded in a capillary (dark blue triangles), and in the Si$_3$N$_4$ membrane.}
  \label{whcomp}
\end{center}
\end{figure}

We also used the Williamson-Hall equation integrated in the \textit{Jana2006} package for verification. This is given by, 
\begin{equation}
\beta \cdot cos(\theta) = \frac{\lambda}{D_v} + \eta \cdot sin(\theta)
\end{equation}
where D$_v$ is the crystallite size and $\eta$ corresponds to the strain which is negligible. A few examples of the sizes found through this method are found in Fig. \ref{whcomp}. The crystallite sizes found using both methods are in close agreement and are shown in Table \ref{tab1}. We find that grain sizes decrease from $\sim$ 8 nm to 4 nm after an exposure of 6.4 $\times$ 10$^{27}$ eV cm$^{-2}$ or 40 minutes of continuous irradiation. However, considering the larger error of the fitted parameters due to the lower S/N ratio of the diffraction peaks (noted in the last column of Table \ref{tab1}) due to loss of crystalline structure, we cannot use the derived crystallite sizes as indicators of the amorphization evolution of the sample. We use other diffractogram parameters, like loss of diffraction peak intensity, peak broadening, and disappearing of discrete spots and arcs, as evidence for the gradual amorphization of this sample. Compared to the beam size of 60 nm, the nanograins do not behave anymore like crystals (the surface effects become dominant), setting some limits in the classical diffraction laws. A note of caution should be taken with the interpretation of the crystallite sizes using the Scherrer equation, a hot topic in the powder diffraction community, e.g. \cite{Scardi2008}. The Scherrer and the Williamson-Hall equations constrain the average crystallite size during irradiation, hardly evolving within the error bars. 

The lattice parameters from the diffractograms are refined versus irradiation exposures and summarized in Table \ref{tab1}. With increasing exposure, the increasing amorphization also meant that the S/N of the diffraction peaks decreased so that the peak positions and crystallite sizes are dominated by systematic uncertainties. Some of the diffractograms could not be adjusted as the S/N was too low and the refinement of the lattice parameters was degenerate. 
Analysis of the same powder sample in the 780 $\mu$m capillary showed the same diffraction peaks and lattice parameters prior to the irradiation series. Diffraction patterns have a higher contrast in the capillary because of the larger thickness ($\phi$ = 780 $\mu$m) compared to the Si$_3$N$_4$ substrate ($\phi$ = 200 $\mu$m). 
The average crystallite size obtained from the Williamson-Hall equation for the sample in the capillary is 75 $\AA$ which is in good agreement with the size found in the Si$_3$N$_4$ substrate. 

\section{Discussion}

X-rays on matter promote ionization, excitation, and a population of secondary photoelectrons. These electrons can cause further ionization and elastic interactions with the target atoms, producing point defects and lattice disorder. 
The deposited energy of a 29.4 keV photon in a single $\phi$ $\sim$ 8 nm enstatite grain (see Section \ref{XRD} for the crystallite size determination before irradiation) is negligible. However for a 200 $\mu$m or 780 $\mu$m column of compacted nanometer sized grains, the transmission at 29.4 keV is $\sim$ 0.953 or 0.829 respectively. Most energy incident on the sample is transmitted by Compton scattered X-rays, with an attenuation length of $\sim$3 mm. Only 5 to 18 \% of the incident energy is absorbed depending on the sample thickness, as fluorescence of O, Mg, Si with attenuation lengths of $\sim$ 0.73, 2.06, 2.41 $\mu$m respectively, and by photoelectrons of energies $\sim$27 - 29 keV with attenuation lengths in enstatite between 4 to 4.4 $\mu$m \citep{Katz1952,Ziaja2006}. 
When considering a column of compacted grains of macroscopic thicknesses (200 or 780 $\mu$m), encased in either Si$_3$N$_4$ or in silica (capillaries), the absorbed X-ray energy is fully deposited within the sample.

The geometry and thermal properties of the sample and sample holder should be considered to fully understand the thermal transfer in the sample. The silica capillary and Si$_3$N$_4$ membrane are themselves producing secondary electrons. The role of the acrylic resin in the Si$_3$N$_4$ membrane samples should also be taken into account. The charge insulation by the resin may further enhance charge accumulation during irradiation. However, previous work showed that classical thermal transfer models may not be enough to understand the heat transfer of nanoparticles as their thermodynamic stability greatly differs from that of the bulk \citep{Kelly2003}.  \cite{Ponomarenko2011} studied radiation-induced melting in X-ray diffractive imaging at the nanoscale by irradiating polymethyl methacrylate (PMMA) samples on a Si$_3$N$_4$ window. They argued that the sizes of their nanoparticles were comparable with characteristic lengths of the heat generation and conduction processes, such as electron or phonon mean free paths. Their calculated rates of temperature rise alone could not explain the melting. Nanometer sized particles may enhance radiation damage as the photoelectron yield is increased \citep{Lewinski2009}. Such electrons have enough energy to escape from one nanoparticle and penetrate into another if the collection is closely packed.

Previous irradiation experiments with silicate analogs (in either amorphous or crystalline form) have mostly involved ion and electron irradiation. Studies with silicate analogs have shown that crystallization of forsterite (MgSiO$_4$) grains is possible via thermal annealing at temperatures between 723 - 873 K \citep{Fabian2000}. \cite{Carrez2002} showed that enstatite grains could be crystallized under 300 keV electron irradiation. 

\cite{Lemelle2003} studied the effect of 30 keV electron irradiation on olivine crystals. In their work, they noted that with increasing exposures, their nanometer sized spherules become larger and more irregular in shape. The highest calculated temperature variation due to heat transfer was 40$^{\circ}$C, which is much below the melting point of olivine (1720$^\circ$C at 1 bar). A simple dissipation of the incident energy into heat could not account for the observed damage. The destabilization of olivine was attributed to electrostatic discharges leading to the breakdown of the dielectric lattice. This occurred when the voltage stress induced by accumulated/trapped charges exceeds the bulk breakdown strength. We interpret the X-ray induced amorphization of our polycrystalline nano-sized enstatite grains embedded in resin, as a result of similar charge build-up processes.

The first ion irradiation experiments of astrophysical interest were performed by  \cite{Day1977} and \cite{Kraetschmer1979}. \cite{Day1977} exposed olivine to 2 MeV protons resulting in no alteration. \cite{Kraetschmer1979} showed that olivine was amorphized following exposure to 1.5 MeV Ne$^+$ ions. Later on, \cite{Bradley1994} showed that irradiation with light H$^+$ and He$^+$ ions at 4 - 20  keV could amorphize the rim around lunar regolith grains and interplanetary dust particles (IDPs). \cite{Demyk2001} showed that the irradiation  with 4 - 10 keV He$^+$ ions destroys the long-range structure in crystalline olivine and later confirmed this for forsterite, enstatite, and diopside grains \citep{Demyk2004}. 
 \cite{Jager2003b} irradiated enstatite grains with 400 keV Ar$^+$ and 50 keV He$^+$ ions. 
 Experiments with 30 - 60 keV H$^+$ ions on forsterite also confirmed amorphization \citep{Brucato2004}.
More recently, \cite{Bringa2007} irradiated forsterite grains  with 10 MeV Xe$^{+3}$ ions. At this energy, 86\% of the total stopping power (energy loss per unit path length) is electronic (4.5 keV nm$^{-1}$), decreasing with penetration depth as the ion slows down. 
Electronic energy loss S$_E$ (or electronic stopping power S$_E$) causes the breaking of bonds, ionization and annealing effects. Nuclear energy loss S$_N$ causes \textit{knock-on} processes and therefore displacements of atoms in the target, which consequently produces amorphization. For heavier ions at similar energy, the fraction of nuclear energy deposition increases \citep{Ziegler2010}. 
Very energetic ions (E $\geq$ 10 MeV) also have low S$_N$/S$_E$ ratio, even though they are heavy.  
With X-rays, the main interactions remains electronic (inducing core electron excitation/ionization), and their effect on matter is more comparable to that of swift heavy ions than lighter ionic species.

\section{Astrophysical implications}

X-rays are generated by many astrophysical sources including massive stars, accreting neutron stars, shocks, supernova, and active galactic nuclei, among others. These X-rays will then interact with the surrounding material composed majorly of gas and dust grains. The launch of the Chandra X-Ray Observatory and the XMM-Newton Observatory in 1999 opened a new era in X-ray astronomy by revealing an abundance of  X-ray dominated regions in the cold and hot universe \citep{Paerels2003}.

The properties of dust grains will depend on their astrophysical environment. Dust grains in the interstellar medium are on average isolated 0.1 $\mu$m sized and quasi-transparent to hard X-rays. Dust grains in protoplanetary disks can grow to cm-size via coagulation by the time the central star becomes optically visible \citep{Natta2007}. In these regions, larger dust grains are able to absorb hard X-rays and undergo alteration. 
Protoplanetary disks are known to have strong X-ray fields arising from the central young stars \citep{Feigelson2003}. \cite{Gorti2009} showed that X-rays play a role on the disk mass-loss rates via the indirect effect of raising the degree of ionization in the disk and increasing the efficiency of FUV-induced grain photoelectric heating, and in cases with high X-ray luminosities, promoting the formation of gaps in the inner r $\sim$10 AU disk.

\cite{Watson2009} studied a large sample of protoplanetary disks in the Taurus-Auriga young cluster and found that, while crystalline silicates are confined to small radii, r $\lesssim$ 10 AU, there is no correlation of the crystalline mass fraction with stellar mass, luminosity, or accretion rate, nor disk factors (mass, disks/star mass ratio). They suggested that X-ray heating may be dominating the heating and annealing of the grains or that another process must be at work within the disks to erase the correlations they produce (like giant planet formation and migration). \cite{Glauser2009} examined the first hypothesis by studying the link between the X-ray luminosity and the X-ray hardness of the central object of 42 T Tauri stars (class II) with the crystalline mass fraction of the circumstellar dust. They found a significant anti-correlation for 20 objects within an age range of $\sim$1 to 4.5 Myr. They determined fluxes around 1 AU and ion energies of the present solar wind are sufficient to amorphize the upper layer of dust grains very efficiently, leading to an observable reduction of the crystalline mass fraction of the circumstellar sub-micron sized dust. This effect could also erase other relations between crystallinity and disk/star parameters such as age or spectral type. More recently, \cite{Cleeves2014} studied the primary ionizing agents in the TW Hya protoplanetary disks: cosmic rays and X-rays. They found that the emission spectra due to HCO$^+$ and N$_2$H$^+$ is best fitted by moderately hard X-rays, dominating over incident cosmic rays. In light of these findings,  two competing mechanisms might be at the origin of the alteration of circumstellar dust. Depending on the age of the central star, X-rays may amorphize grains efficiently, while thermal annealing near the star will crystallize dust grains. Although grain growth could play a role in the disappearance of crystalline features in the mid-infrared, the coagulation of crystalline grains into aggregates is possible. Crystalline silicates observed in disks are significantly smaller than amorphous grains \citep{Juhasz2010} and could assemble into larger amorphous aggregates \citep{Watson2009, Merin2007}. In this way, crystalline silicate bands will appear as if the crystals were isolated \citep{Min2008}. X-rays can transform the crystals in the aggregates into amorphous grains so that crystalline spectral features are lost.  Further observations from the mid-infrared to the millimeter, models including grain size distributions, and laboratory experiments, are needed to distinguish between grain growth/coagulation and amorphization/annealing in protoplanetary disks.

From all our X-ray irradiated samples, only the crystalline enstatite grains embedded in cyanoacrylate, an organic resin, underwent amorphization. The free crystalline grains in the capillary (without resin) did not undergo amorphization. In addition, none of the amorphous silicate samples evolved: neither in the resin-embedded Si box nor in the capillary holder. We interpret this by the fact that free sub-micron sized grains are quasi-transparent to hard X-rays. Dust grains in the interstellar medium, where grains are sub-micron sized and isolated, will likely not absorb hard X-rays as long as they are free. On the other hand, the sample that did undergo amorphization was embedded in an organic matrix, acting as a coagulation source, increasing the volume of the sample and allowing charge-build up by secondary photoelectrons. 
More experiments with different organic materials will be required to further explore the role of organics in coagulation and charge build-up under irradiation. Organic/mineral admixtures are present in space, as shown by an study of returned IDPs by \cite{Bradley2005}. They showed that GEMS contain amorphous silicates surrounded by carbon matter (a mixture of inorganic and organic carbon), with isotopic ratios signaling its interstellar origin. Indeed, during the passage from cold dark clouds to protoplanetary disks, molecules freeze out from the gas phase onto dust grain surfaces, producing icy mantles \citep{Bergin2007}. Under irradiation (photons/ions/electrons) these icy mantles can produce organic residues on mineral grain surfaces. These organic residues can form in molecular clouds, but also in protoplanetary disks and protostellar cores. Protoplanetary disks contain silicates but also carbon dust  \citep{Lisse2006, Tielens2008} and ices \citep{Carr2008}. Organic solids can play a role in the amorphization of silicate-organic aggregates in the X-ray dominated regions of protoplanetary disks via charging of the particles upon X-ray irradiation.
 
\section{Conclusions}

Hard X-rays are present in several astrophysical environments.  The goal of this study is to simulate the exposure of cosmic dust to hard X-rays and understand the potential effects on the dust grain structure. In order to do this, we prepared cosmic dust grain analogs: crystalline and amorphous MgSiO$_3$ grains, with an average size of $\sim$8 nm. We irradiated these grains with hard X-ray nanobeams produced at the European Synchrotron Radiation Facility (Grenoble), at beamline ID16B-NA, and followed their evolution in-situ via X-ray diffraction. While irradiation with the monochromatic beam at 29.4 keV gives an upper limit to the cumulative effect of hard X-rays, it does not simulate realistic stellar fluxes in these astrophysical environments. Future experiments are planned with broad spectrum and soft X-ray sources that will complement this work.

An evolution of the structure of crystalline grains embedded in an organic resin towards amorphization was detected after analysis of their diffractograms. The net amorphization of the grains was attributed to charging effects enhanced by the organic matrix, increasing the effective volume of the sample. The reverse process, crystallization of amorphous silicates, was not detected for equivalent X-ray exposures. None of the free samples (without cyanoacrylate) showed any alteration, a result applicable to the interaction of hard X-rays with isolated sub-micron dust grains in the interstellar medium. Our results manifest the capacity of hard X-rays to modify the structure of agglomerated nanoparticles in higher density regions irradiated by young stars such as protoplanetary disks, where grains can coagulate into cm-sizes and beyond. The results of these high exposure X-ray irradiation experiments have important consequences for the analysis of microscopic mineral samples with synchrotron radiation. In particular, the use of highly insulating materials such as aerogel could impact the analysis of curated dust return samples. High-fluences would then have to be carefully limited.  

\section{Acknowledgments}

We acknowledge the European Synchrotron Radiation Facility for provision of synchrotron radiation facilities (proposal No. HC-1709) and we thank the beamline scientists of ID16B. We thank Gabriele Born (Astrophysikalisches Institut Jena) for her assistance in sample preparation.

\bibliography{xray}

\end{document}